# REVISION OF THE PHENOMENOLOGICAL CHARACTERISTICS OF THE ALGOL-TYPE STARS USING THE NAV ALGORITHM

M.G. Tkachenko[1], I.L. Andronov[1], L.L. Chinarova[2]

[1] Department "High and Applied Mathematics", Odessa National Maritime University,
Mechnikova st., 34, 65029, Odessa, Ukraine,
*masha.vodn@yandex.ua, tt_ari@ukr.net*

[2] Astronomical Observatory, Odessa National University,
Shevchenko Park, 65014, Odessa, Ukraine, *lidia_chinarova@mail.ru*

ABSTRACT. Phenomenological characteristics of the sample of the Algol – type stars are revised using a recently developed NAV ("New Algol Variable") algorithm (2012Ap.....55..536A, 2012arXiv 1212.6707A) and compared to that obtained using common methods of Trigonometric Polynomial Fit (TP) or local Algebraic Polynomial (A) fit of a fixed or (alternately) statistically optimal degree (1994OAP.....7...49A, 2003ASPC..292..391A).

The computer program NAV is introduced, which allows to determine the best fit with 7 "linear" and 5 "non-linear" parameters and their error estimates. The number of parameters is much smaller than for the TP fit (typically 20 – 40, depending on the width of the eclipse, and is much smaller (5 – 20) for the W UMa and β Lyrae – type stars. This causes more smooth approximation taking into account the reflection and ellipsoidal effects (TP2) and generally different shapes of the primary and secondary eclipses. An application of the method to two – color CCD photometry to the recently discovered eclipsing variable 2MASS J18024395 + 4003309 = VSX J180243.9 +400331 (2015JASS...32..101A) allowed to make estimates of the physical parameters of the binary system based on the phenomenological parameters of the light curve. The phenomenological parameters of the light curves were determined for the sample of newly discovered EA and EW – type stars (VSX J223429.3+552903, VSX J223421.4+553013, VSX J223416.2+553424, USNO-B1.0 1347-0483658, UCAC3-191-085589, VSX J180755.6+074711= UCAC3 196-166827). Despite we have used original observations published by the discoverers, the accuracy estimates of the period using the NAV method are typically better than the original ones.

## 1. Introduction

Phenomenological modeling is an effective tools to study newly discovered or poorly studied eclipsing binary stars, for which there is no sufficient information on spectra/temperatures and mass ratio. This additional information is needed for physical modeling using the algorithm of Wilson and Devinney (1971) also discussed by Wilson (1994). This method is implemented in some different programs, like BinaryMaker (Bradstreet, 2005), PHOEBE (Prsa A. et al., 2011), series of programs written in the Fortran language (Zoła et al. 1997, 2010).

The physical modeling of close binary stars is discussed in detail by Kopal (1957), Tsessevich (1971), Kallrath and Milone (2009).

The phenomenological modeling of variable stars has a long history starting from hand-drawing of the light curve and further algebraic polynomial fits of the parts of the light curve or trigonometric polynomial (TP) fits.

The statistically optimal degree of the polynomial (or other more complicated model) may be determined using at least three criteria and their modifications (see Andronov 1994, 2003 for reviews).

## 2. Approximations of Minima

The "local" approximations of the extrema including polynomials, splines, asymptotic parabolae, asymmetric hyperbolae, running parabolae and sines were discussed by Andronov (2005). The Gaussian shape was an usual approximation for spectral lines in the era preceding synthetic spectra, and, besides measuring Doppler shifts, was used also for minima determination of eclipsing variables and, in a modified form, by Mikulášek et al. (2012). However, the Gaussian function has no abrupt switch from zero to the zero profile, thus it is not possible to determine from the fit the value of the eclipse duration – one of the main parameters needed for the General Catalogue of Variable Stars (Samus' et al., 2015).

Andronov (2012ab) compared few limited – width approximations and proposed the following approximation:

$$x_C = C_1 + C_2 \cos(2\pi\phi) + C_3 \sin(2\pi\phi) + \\ + C_4 \cos(4\pi\phi) + C_5 \sin(4\pi\phi) + \\ + C_6 H(\phi/C_8, C_9) + C_7 H((\phi-0.5)/C_8, C_{10}) \quad (1)$$

Here the shape of the eclipse is described as

$$H(z,\alpha) = \begin{cases} (1-|z|^\alpha)^{3/2}, & \text{if } |z| \leq 1 \\ 0, & \text{if } |z| > 1 \end{cases} \quad (2)$$

The usual determination of phase is $\phi = \zeta - \text{int}(\zeta - \zeta_0)$, where $\zeta = (t - T_{00})/P_0$, $T_{00}$ is the initial epoch, $P_0$ is period, and $\zeta_0$ is the minimal limit of the interval ($\zeta_0 \leq \phi < \zeta_0 + 1$) of possible values of $\phi$. Classical approach is the interval of phases $[0,1)$, i.e. $\zeta_0 = 0$, with an additional notification that one may extend the main interval by adding any integer number $E$ to the phase, as the light curve is suggested to be periodic: $m(\phi + E) = m(\phi)$. This obvious extension may be realized at computer programs either by doubling the data with a shift of $E = 1$, or by using suitable values of $\zeta_0$ for different parts of the light curve. As typically the initial epoch and period are defined so that the primary minimum corresponds to $\phi = 0$ (or close to 0) and the secondary minimum – to $\phi = 0.5$. Thus for Eq. (1) it is suitable to choose $\zeta_0 = -0.25$.

In previous papers, we have used the values $T_{00}$ and $P_0$, which were determined by other methods (e.g. trigonometric polynomial). Andronov et al. (2015b) improved the method NAV to make possible differential corrections

$$\phi = \zeta - \text{int}(\zeta - \zeta_0) + C_{11} + C_{12} \cdot (t - T_1), \quad (3)$$

where $T_1$ is some moment of time, which is recommended to be close to the mean time $\bar{t}$ for partial orthogonalization of basic functions (Andronov, 1994, 2003). The corrected values of the light elements are (Andronov et al. 2015b):

$$T_{01} = \frac{T_{00} + C_{11}P_0 - C_{12}P_0 T_1}{1 - C_{12}P_0} \quad (4)$$

$$P_1 = \frac{P_0}{1 - C_{12}P_0} \quad (5)$$

We also use additional parameters (the relative depths of the primary and secondary minima, respectively) $d_1 = 1 - 10^{-0.4C_6}$, $d_2 = 1 - 10^{-0.4C_7}$ and their combinations $Y = d_1 + d_2$ and $\xi = d_1/d_2 = F_1/F_2$. Here $F_1$ and $F_2$ are relative values of the mean brightness of the eclipsed part of star. The value of $Y$ varies from 0 (no eclipses) to 1 (both full eclipses).

Assuming a simplified model of uniform brightness distribution and spherical symmetry of components (Tsessevich 1971, Shulberg 1971, Andronov 1991, Malkov et al. 2007, Andronov and Tkachenko 2013), one may estimate physical parameters

As generally the primary minimum is defined to be more deep than the secondary one (however, it may be not the case, if the minima are of the same depth within error estimates), it corresponds to the case, when the star with larger surface brightness (and so temperature) is eclipsed by a cooler component.

Papageorgiou et al. (2014) proposed the simplest parabolic approximation of the light curve in four fixed phase intervals, which has the simplest program realization. The number of parameters (12) is the same as in our NAV approximation, but the NAV approximations are continuous, allow to determine the width and corrections to the initial epoch and period. However, the NAV approximation needs more computational time.

Mikulášek (2015) proposed another special shape for the eclipse as

$$F_e = A\left(1 + C\frac{\phi^2}{D^2} + K\frac{\phi^4}{D^4}\right)\left\{1 - \left\{1 - \exp\left[1 - \cosh\left(\frac{\phi}{D}\right)\right]\right\}^r\right\} \quad (6)$$

from which some are similar to that we used in the NAV approximation, i.e. the characteristic width $D$ is proportional to the eclipse half-width $C_8$, $A = C_6$ (for a primary minimum) or $A = C_7$ (for a secondary minimum), and, for small phases $\phi \to 0$, $\alpha \to 2r$ for "sharp" eclipses $r \leq 1$ and else $\alpha \to 2$. Two additional parameters $C$, $K$ should generally produce closer coincidence of the approximation to the observations, but also larger statistical errors of the parameters. Obviously, the number of parameters may also be increased in the NAV algorithm, as was discussed by Andronov (2012ab), but currently we try to make studies using small modifications of the initial algorithm.

## 3. Application to Concrete Stars

For the analysis, we have used photometric observations of 6 newly discovered eclipsing binary stars of different types, which were published in the "Open European Journal on Variable Stars" (OEJV).

The phase light curves and their best fit appoximations (including non-linear optimization for the parameters $C_8$ …$C_{12}$) are shown in Fig. 1 – Fig. 6. For the preparation of figures, the program MCV (Andronov and Baklanov 2004) was used. The parameters are listed in Table 1.

The analysis shows that the corrections of the period related to the parameter $C_{12}$ range typically from $1\sigma$ to $7\sigma$, so sometimes are statistically significant. The error estimates for the period using our method are typically better than the original ones, for two stars from 6 by a factor of 10-12. Similar situation is present for the initial epochs. We conclude that the current improvement of the method by adding parameters $C_{11}$ and $C_{12}$ is efficient for the determination of the initial epoch and the photometric period.

Among the coefficients $C_2$ …$C_5$ for this sample of stars, the largest value for a given star is typically of $C_4$, which corresponds to a "double frequency wave" due to the ellipticity effect.

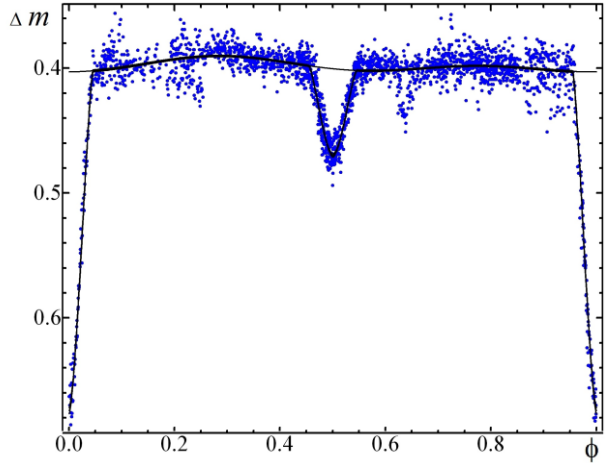

Figure 1: The phase light curve for the star VSXJ223421.4+553013. The observations are shown as blue dots, the black lines show the NAV approximation and ±1σ error corridors.

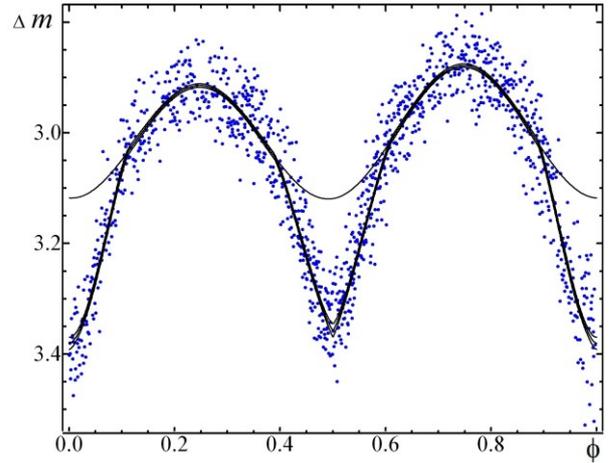

Figure 4: The phase light curve for the star USNO-B1.0 1347-0483658. The observations are shown as blue dots, the black lines show the NAV approximation and ±1σ error corridors.

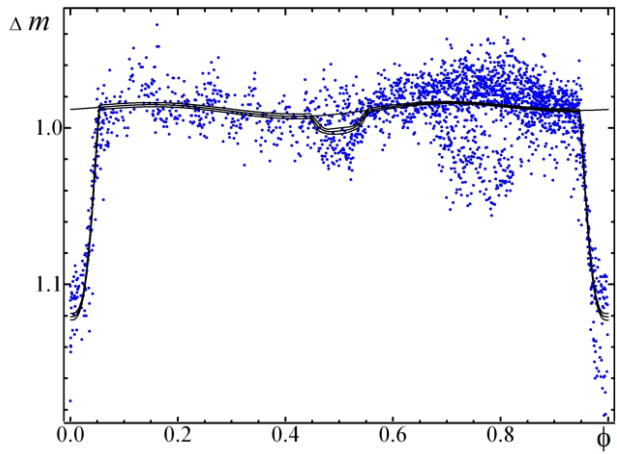

Figure 2: The phase light curve for the star VSXJ223416.2+553424. The observations are shown as blue dots, the black lines show the NAV approximation and ±1σ error corridors.

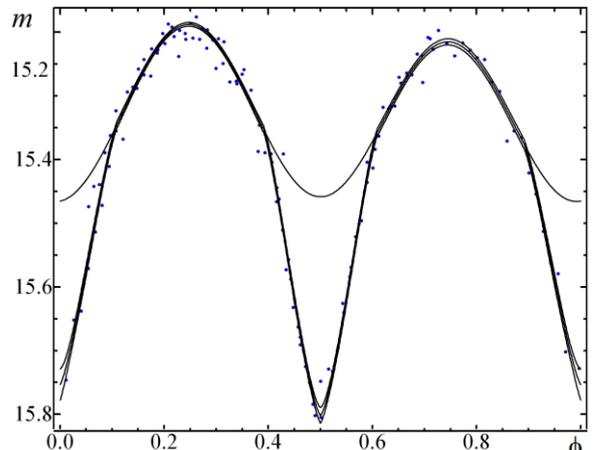

Figure 5: The phase light curve for the star UCAC3 191-085589. The observations are shown as blue dots, the black lines show the NAV approximation and ±1σ error corridors.

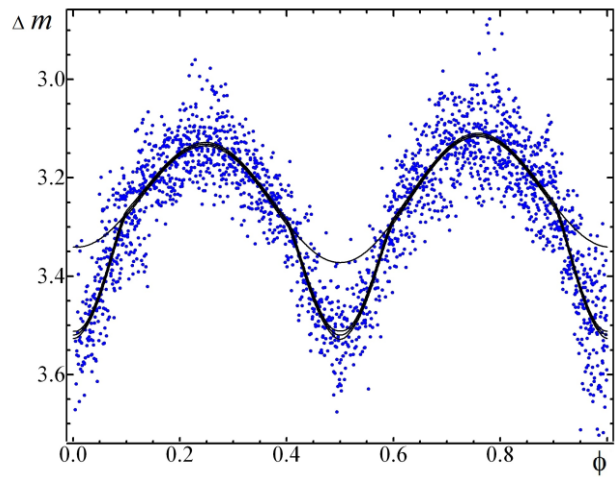

Figure 3: The phase light curve for the star VSXJ223429.3+552903. The observations are shown as blue dots, the black lines show the NAV approximation and ±1σ error corridors.

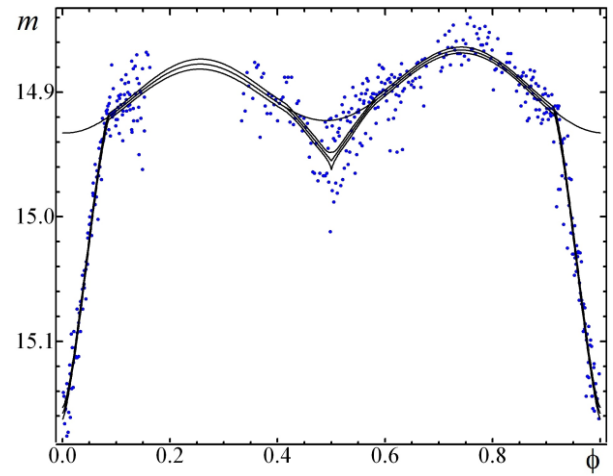

Figure 6: The phase light curve for the star UCAC3 196-166827=VSX J180755.6+074711. The observations are shown as blue dots, the black lines show the NAV approximation and ±1σ error corridors.

Table 1. Characteristics of the NAV approximations

| | VSX J223421.4+553013 (Lehký 2009a, OEJV 99, #3) | VSX J223416.2+553424 (Lehký 2009a, , OEJV 99, #2) | VSX J223429.3+552903 (Lehký 2009a, OEJV 99, #1) |
|---|---|---|---|
| $C_1$ | 0.3980±0.0003 | 0.9876±0.0006 | 3.2393± 0.0027 |
| $C_2$ | 0.0013±0.0004 | -0.0014±0.0009 | -0.0159±0.0032 |
| $C_3$ | -0.0038±0.0004 | 0.0014±0.0007 | 0.0092±0.0020 |
| $C_4$ | 0.0036±0.0004 | 0.0019±0.0007 | 0.1170±0.042 |
| $C_5$ | 0.0015±0.0004 | -0.0024±0.0007 | 0.0040±0.0038 |
| $C_6$ | 0.2714±0.0026 | 0.1325±0.0025 | 0.1786±0.0103 |
| $C_7$ | 0.0695±0.0015 | 0.0109±0.0023 | 0.1475±0.0109 |
| $C_8$ | 0.0456±0.0004 | 0.0546±0.0008 | 0.0985±0.0042 |
| $C_9$ | 1.56±0.04 | 3.19±0.19 | 1.93±0.24 |
| $C_{10}$ | 1.69±0.08 | 3.97±2.33 | 1.96±0.30 |
| $C_{11}$ | -0.0002±0.0001 | 0.0010±0.0003 | -0.0067±0.0014 |
| $C_{12}$ | (-1.89±0.93) ·$10^{-6}$ | (4.08±1.93) ·$10^{-6}$ | (-3.35±0.50) ·$10^{-5}$ |
| $T_{00}$ | 54373.28108±0.00035 | 54387.58563±0.00040 | 54387.35274±0.00040 |
| $P_0$ | 1.324325±0.000150 | 1.105815±0.000009 | 0.387245±0.000009 |
| $T_{01}$ | 54442.14572±0.00017 | 54466.09957±0.00035 | 54462.47568±0.00056 |
| $P_1$ | 1.3243217±0.0000012 | 1.1058209±0.0000021 | 0.3872400±0.0000020 |
| $d_1$ | 0.2212±0.0018 | 0.1148±0.0020 | 0.1517±0.0080 |
| $d_2$ | 0.0620±0.0013 | 0.0100±0.0021 | 0.1270±0.0087 |
| $Y$ | 0.2832±0.0024 | 0.1248±0.0030 | 0.2787±0.0140 |
| $\gamma$ | 3.5662±0.0759 | 11.4929±2.4366 | 1.1941±0.08260 |
| Min I | 0.674 ±0.002 | 1.121±0.002 | 3.519 ±0.007 |
| Max I | 0.390 ±0.001 | 0.984±0.001 | 3.112 ±0.003 |
| | USNO-B1.0 1347-0483658 (Lehký 2009b, OEJV 115, #1) | UCAC3 191-085589 (Moos et al. 2013, OEJV 156, #6) | VSX J180755.6+074711 (Franco et al. 2010, OEJV 135, #1) |
| $C_1$ | 3.0073±0.0033 | 15.3314±0.0035 | 14.8996±0.0012 |
| $C_2$ | -0.0005±0.0033 | 0.0034±0.0040 | 0.0050±0.0017 |
| $C_3$ | 0.0180±0.0018 | -0.0139±0.0025 | 0.0056±0.0018 |
| $C_4$ | 0.1109±0.0047 | 0.1301±0.0051 | 0.0278±0.0026 |
| $C_5$ | -0.0035±0.0034 | -0.0066±0.0038 | -0.0006±0.0014 |
| $C_6$ | 0.2627±0.0142 | 0.2882±0.0256 | 0.2250±0.0061 |
| $C_7$ | 0.2383±0.0140 | 0.3432±0.0153 | 0.0324±0.0071 |
| $C_8$ | 0.1124±0.0042 | 0.1095±0.0037 | 0.0866±0.0023 |
| $C_9$ | 1.65±0.18 | 1.24±0.18 | 1.41±0.08 |
| $C_{10}$ | 1.23±0.13 | 1.38±0.11 | 0.41±0.63 |
| $C_{11}$ | 0.0012±0.0012 | 0.0022±0.0011 | -0.0011±0.0007 |
| $C_{12}$ | (3.64±3.94) ·$10^{-6}$ | 0.0048±0.0014 | 0.0004±0.0002 |
| $T_{00}$ | 55068.50679± 0.00075 | 55948.67692±0.00405 | 55381.44874±0.00344 |
| $P_0$ | 0.2576355± 0.0000009 | 0.27448±0.00107 | 0.861209±0.00013 |
| $T_{01}$ | 54881.46373± 0.00030 | 55949.22649±0.00029 | 55387.47629±0.00058 |
| $P_1$ | 0.2576357±0.0000010 | 0.27484 ±0.00038 | 0.86154±0.00014 |
| $d_1$ | 0.2149±0.0103 | 0.2331±0.0181 | 0.1879±0.0055 |
| $d_2$ | 0.1971±0.0104 | 0.2710±0.0103 | 0.0294±0.0064 |
| $Y$ | 0.4120±0.0171 | 0.5041±0.0227 | 0.2165±0.0083 |
| $\gamma$ | 1.0906±0.0613 | 0.8602±0.0674 | 6.3740±1.3713 |
| Min I | 3.380± 0.011 | 15.753±0.025 | 15.157±0,004 |
| Max I | 2.878± 0.003 | 15.187±0.003 | 14.866±0.002 |

The stars USNO-B1.0 1347-0483658 and UCAC3 191-085589 show unequal maxima (O'Connell (1951) effect). This effect is more pronounced in the coefficient $C_3$ than $C_5$. This is in a good agreement with theoretical expectations (cf. Davidge and Milone, 1984).

In future, possibly it will have sense to exclude the term with $C_5$ from the mathematical model (1), if still being within the error estimates. This will be decided after analyzing a larger sample of stars.

The parameter $\alpha$ (Eq. (2)) (i.e. $C_9$, $C_{10}$ for the primary and secondary minimum, respectively) is typically seen in the range from 1 to 2. The outstandingly small value of $C_{10}$=0.41±0.63 for VSX J180755.6+074711 also does not differ from this range within error estimates.

In our sample, the exception for both minima is present in VSX J223416.2+553424. The secondary eclipse is very shallow, but is still seen in the noisy observations,

Large values $\alpha$>>2 typically correspond to a total eclipse or a transit and a large difference in radii. For this case, Andronov (2012a) proposed a possible extension for the expression (2) assuming that the eclipse starts at some non-zero value $z_0$, so

$$0 \le z_0 \le z \le 1. \quad (7)$$

For the present noisy observations, such a complication due to using an additional parameter $z_0$ seems not to be effective. This object is also extreme in our sample due to a very large brightness ratio γ=11.5±2.5 and thus large relative temperature difference. Multi-color are needed for a more detailed study similar to that described by Andronov et al. (2015a) for the object 2MASS J18024395 + 4003309 = VSX J180243.9 +400331.

### 4. Conclusions

Parameters of the phenomenological modeling are determined for 6 newly discovered eclipsing binary stars of different types.

The initial algorithm (Andronov 2012ab) was improved by adding two parameters allowing to make corrections to the initial epoch and period. The corresponding error estimates are typically significantly better than that obtained using other methods. This improvement of accuracy is more efficient for the EA – type stars, i.e. relatively short eclipses.

In our sample, there are 3 stars of the W UMa – type. They all show distinct eclipses, contrary to the GCVS statement that EB and EW – type objects are systems "having light curves for which it is impossible to specify the exact times of onset and end of eclipses" (Samus' et al. 2015). We propose to correct this definition.

The presence of eclipses is to be justified by statistical significance of the parameters $C_6$ and $C_7$ (eclipse depth), otherwise should be classified as "elliptic" (ELL).

For a fixed declination $i$, one may expect an increase with relative radii of the components (and thus from EA to EW types) of the parameters $C_8$ (eclipse half-width), $C_2$, $C_4$ (proximity effects), $C_6$ and $C_7$ (eclipse depth).

*Acknowledgements.* The authors are thankful to Professors S. Zoła and Z. Mikulášek for fruitful discussions over the years. This research is a part of the projects "Inter – Longitude Astronomy" (Andronov et al. 2010) and "Ukrainian Virtual Observatory" (Vavilova et al. 2012).